# Superconductors, Orbital Magnets, and Correlated States in Magic Angle Bilayer Graphene

Xiaobo Lu[1], Petr Stepanov[1], Wei Yang[1], Ming Xie[2], Mohammed Ali Aamir[1], Ipsita Das[1], Carles Urgell[1], Kenji Watanabe[3], Takashi Taniguchi[3], Guangyu Zhang[4], Adrian Bachtold[1], Allan H. MacDonald[2] and Dmitri K. Efetov[1]*

1. ICFO - Institut de Ciencies Fotoniques, The Barcelona Institute of Science and Technology, Castelldefels, Barcelona, 08860, Spain
2. Department of Physics, University of Texas at Austin, Austin TX 78712, USA
3. National Institute for Materials Science, 1-1 Namiki, Tsukuba, 305-0044, Japan
4. Beijing National Laboratory for Condensed Matter Physics and Institute of Physics, Chinese Academy of Sciences, Beijing, 100190,China

*E-mail: dmitri.efetov@icfo.eu

**Superconductivity often occurs close to broken-symmetry parent states and is especially common in doped magnetic insulators[1]. When twisted close to a magic relative orientation angle near 1°, bilayer graphene has flat moiré superlattice minibands that have emerged as a rich and highly tunable source of strong correlation physics[2-5], notably the appearance of superconductivity close to interaction-induced insulating states. Here we report on the fabrication of bilayer graphene devices with exceptionally uniform twist angles. We show that the reduction in twist angle disorder reveals insulating states at all integer occupancies of the four-fold spin/valley degenerate flat conduction and valence bands, i.e. at moiré band filling factors $\nu = 0, \pm 1, \pm 2, \pm 3$, and superconductivity below critical temperatures as high as ~ 3 K close to $-2$ filling. We also observe three new superconducting domes at much lower temperatures close to the $\nu = 0$ and $\nu = \pm 1$ insulating states. Interestingly, at $\nu = \pm 1$ we find states with non-zero Chern numbers. For $\nu = -1$ the insulating state exhibits a sharp hysteretic resistance enhancement when a perpendicular magnetic field above 3.6 tesla is applied, consistent with a field driven phase transition. Our study shows that symmetry-broken states, interaction driven insulators, and superconducting domes are common across the entire moiré flat bands, including near charge neutrality.**

Interactions dominate over single-particle physics in flat-band electronic systems, and give rise to insulating states at partial band fillings[3,4,6], superconductivity[7-9], magnetism[6,10-16], charge density waves[17], quasiparticles with exotic statistics[18-22] and other phenomena[23-28]. The recent discovery of correlated insulating phases and strongly coupled superconducting domes in the ultra-flat bands of magic angle twisted bilayer graphene (MAG) close to half-filling ($\nu = \pm 2$)[3,29,30] has established graphene as a new and exciting platform for the investigation of strongly-correlated two-dimensional electrons. MAG promises to provide entirely new insights because correlations can be accurately controlled by varying twist angle, because techniques for the fabrication of ultra-clean graphene layers have been perfected, and because the electron density $n = n_0 = A_0^{-1} \sim 10^{12}$ cm$^{-2}$ needed to fill a moiré superlattice band can be supplied by electrical gates ($A_0$ is the unit cell area of the periodic moiré pattern).

Here we report the observation of correlated states at all integer fillings of $\nu = n/n_0$, including at charge neutrality, and the occurrence of novel superconducting domes upon doping slightly away from these densities. When interactions are neglected, the two (conduction and valence) low-energy moiré bands of MAG have 4-fold spin/valley flavor degeneracies, implying that the density measured from the carrier neutrality point (CNP) is $4n_0$ when the flat conduction band is full and $-4n_0$ when the valence band is empty[2,31]. Interactions can lift the flavor degeneracies and give rise to completely empty or full spin/valley polarized flat bands, with interaction induced gaps at all integer values of $\nu$, in place of the symmetry protected Dirac points that connect the conduction and valence bands of fore each flavor[10] when interactions are neglected. The many-body physics of these bands are intricately sensitive on the twist-angle $\theta$ and interaction strength $\varepsilon^{-1}$ (where $\varepsilon$ is the effective dielectric constant in MAG).

Depending on electronic structure details, bands can have non-zero Chern numbers[10,11,32,33], allowing for the possibility of orbital magnetism and anomalous Hall effects. Gapped states at non-zero $\nu$ occur only when interactions are strong enough to shift band energies by more than the flat band width when they are occupied, otherwise they lead to semi-metallic states.

## Correlated states at all integer moiré filling factors

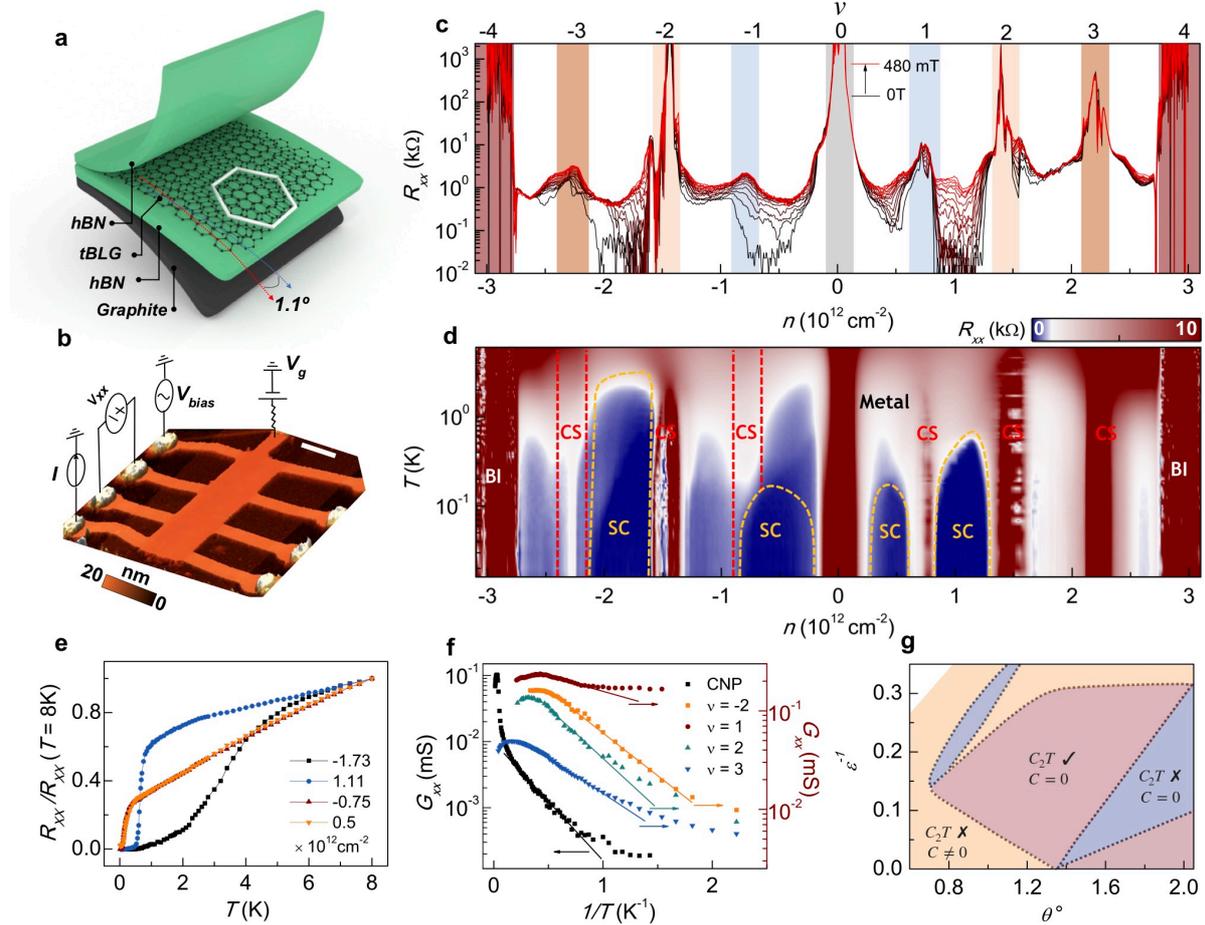

**Figure 1 | Integer-filling correlated states and new superconducting domes. a,** Schematic of a typical hBN encapsulated MAG device with a graphite back gate. **b,** AFM image and four-probe measurement schematic, with the scale bar 2 μm. **c,** 4-terminal longitudinal resistance $R_{xx}$ as a function of carrier density $n$ at different perpendicular magnetic fields from 0T (black trace) to 480mT (red trace). **d,** Color plot of $R_{xx}$ vs. $n$ and $T$, showing different phases including metal, band insulator (BI), correlated state (CS) and superconducting state (SC). The boundaries of the superconducting domes indicated by yellow lines are defined by 50% resistance values relative to the normal state. Note that the metal-SC transition is not sharp at some carrier densities, adding uncertainty to the $T_c$ extraction **e,** Longitudinal resistance $R_{xx}$ at optimal doping of the superconducting domes as a function of temperature. The resistance is normalized to its value at 8K. **f,** Conductance $G_{xx}$ vs. inverse temperature at $n$ corresponding to $\nu = 0, 1, \pm 2$ and 3. The straight lines are fits to $\sim exp(-\Delta/2kT)$ activated behavior and give gap values of 0.35 meV ($\nu = -2$), 0.14 meV ($\nu = 1$), 0.37 meV ($\nu = 2$), 0.27 meV ($\nu = 3$) and 0.86 meV (CNP/ $\nu = 0$). **g,** Mean-field phase diagram for neutral $\nu = 0$ (CNP) twisted bilayer graphene, as a function of twist angle $\theta$ and interaction strength $\varepsilon^{-1}$, showing differnet configurations of $C_2T$ symmetry and Chern number ($C$).

Fig. 1a shows the typical device schematic of a graphite back-gated, hexagonal boron nitride (hBN) encapsulated MAG hetero-structure. Our stack was fabricated using a previously developed "tear and stack" technique[34,35], followed by a mechanical squeezing process[36]. This process removes trapped blisters, releases local strain, and achieves more homogenous interfaces between the layers. The stack was then etched and edge contacted[37] to form a multi-terminal transport device. The atomic force microscopy (AFM) image in Fig. 1b demonstrates the high structural homogeneity of the final device.

Fig. 1c shows the 4-terminal longitudinal resistance $R_{xx}$(n) as a function of carrier density for different values of weak out-of-plane magnetic field $B_\perp$ at base temperature of 16mK. Here $n$ is capacitively tuned by a voltage on the local graphite back-gate and normalized by Hall measurement (Extended Data Fig.1). We find strong resistance peaks at density values $n = 4n_0 \sim \pm 3 \times 10^{12}$ cm$^{-2}$ that mark the edges of the flat moiré bands and are consistent with previous studies[3,29,30]. This full-band density corresponds to an average twist angle across the device $\theta \sim 1.10°$. Comparing $4n_0$ values extracted from 2-terminal measurements between different contact pairs (Extended data Fig.2), we estimate that the twist angle variation is only $\Delta\theta < 0.02°$ over a span of ~10 μm, demonstrating unprecedented twist-angle homogeneity in a MAG device[30].

In addition to resistance peaks at the CNP and $\nu = \pm 4$, we also see resistance peaks at all non-zero integer fillings of the moiré conduction and valence bands ($\nu = \pm 1, \pm 2, \pm 3$) corresponding to 1, 2 and 3 electron (holes) per moiré unit cell (Fig. 1c). Insulating states at these densities are not expected in a single-particle picture, and can only be explained by strong interactions. Signatures of most of these resistive states were previously observed[3,10,29,30,38], but are more strongly developed here, showing higher resistance values at base temperature. In particular, we observe a new resistance peak at $-n_0$ filling, and for the first time resolve resistive states at all integer moiré filling factors of both conduction and valence moiré bands. While the conductance at $\nu = -3$ and $\nu = -1$ does not yet show activated temperature dependence, which might indicate correlated semi-metal states[39,40], all other integer filling factor states exhibit activated behavior over a decade in temperature, as shown in Fig. 1f. The transport data are consistent with gapped or semi-metallic correlated insulator states, with extracted gap values of 0.34 meV ($\nu = -2$), 0.14 meV ($\nu = 1$), 0.37 meV ($\nu = 2$) and 0.25 meV ($\nu = 3$).

Unlike previous work, our device shows clear activated temperature dependence below 33 K at the CNP, with an extracted gap size of 0.86 meV. Gaps at the CNP do not require broken flavor symmetries, but they do require that at least one of the emergent $C_3$ and $C_2T$ symmetries that protect the CNP be broken[11]. Interestingly, because the CNP gap was not observed in more inhomogeneous devices, our findings suggest that the gapped state at the CNP is less robust against twist angle disorder than the non-zero integer filling factor states, possibly because the competing gapless semiconductor single-particle state already has a soft gap. We do note that a gap at the CNP can be also induced by inversion-symmetry breaking by nearly aligned hBN layers[41-43]. While we cannot completely rule out this trivial single particle mechanism as the source of the CNP gap, the hBN alignment in our device was random and we do not observe other typical signatures of hBN alignment[41,42,44-47].

The existence of a gap at the CNP has strong implications for the nature of other gapped MAG states. The mean-field theory (Extended data L-M) phase diagram in Fig.1g suggests that the interacting state is gapped at all plotted twist angles $\theta$ and $\varepsilon^{-1}$ values. Red color indicates gapped states that do not break $C_2T$ symmetry, and therefore have bands with no Berry curvature and vanishing Chern numbers. Blue zones signify gapped states that break $C_2T$ symmetry but nevertheless have bands with zero Chern number, while orange zones indicate bands with non-zero Chern numbers. Gapped states at non-zero integer values of $\nu$ are expected only when the moiré superlattice band width is smaller than the exchange-shift produced by band occupation, and this occurs only near the magic angles. Overall our calculation demonstrate that insulating, or for weak interactions semi-metallic states, are common at all integer values of $\nu$, as observed experimentally.

## Superconducting domes in proximity to $\nu = 0$ and $\nu = \pm 1$

Strikingly, the resistivity of our device measured at 16 mK drops to zero over four distinct intervals of fractional carrier density, and is restored to normal values by a small perpendicular magnetic field $B_\perp <$ 500 mT (Extended Data Fig.3). Fig. 1d provides a color scale plot of resistance versus temperature and carrier density in which red regions correspond to high and dark blue regions to low resistance values

(<50 Ω). Extended Data Fig. 4 displays the resistance versus carrier density at a series of temperatures. At the lowest temperature, dome shaped pockets of low resistance flank the most resistive states. As shown in the resistance versus temperature line-cuts at optimal doping of the domes in Fig. 1e, in four of these regions the resistance drops sharply to almost zero values, consistent with a superconducting phase transition.

The maximum superconducting transition temperature $T_c$ in the superconducting dome (defined by half normal state resistance) on the hole side of $-2n_0$, previously observed as ~1.7 K[29], is increased to $T_c >$ 3 K in our device, with resistive transition widths $T_w \sim 2$ K. All other superconducting domes are observed here for the first time. We identify domes between the CNP and $\pm n_0$, with lower $T_c$ and sharper resistive transitions with $T_c \sim 160$ mK and $T_w \sim 50$ mK for the electron side and $T_c \sim 140$ mK and $T_w \sim 60$ mK on the hole side. In addition, we find a superconducting dome between $n_0$ and $2n_0$ with $T_c \sim 650$ mK and $T_w \sim 100$ mK. The magneto-resistance data in Fig.1c and in Extended data Fig. 5 suggest that additional superconducting domes are likely developing between $\nu = -4$ and $\nu = -3$, $\nu = -2$ and $\nu = -1$, $\nu = 2$ and $\nu = 3$, and possibly between $\nu = 3$ and $\nu = 4$ may be developing, but are obscured by the inhomogeneity that remains in our improved samples.

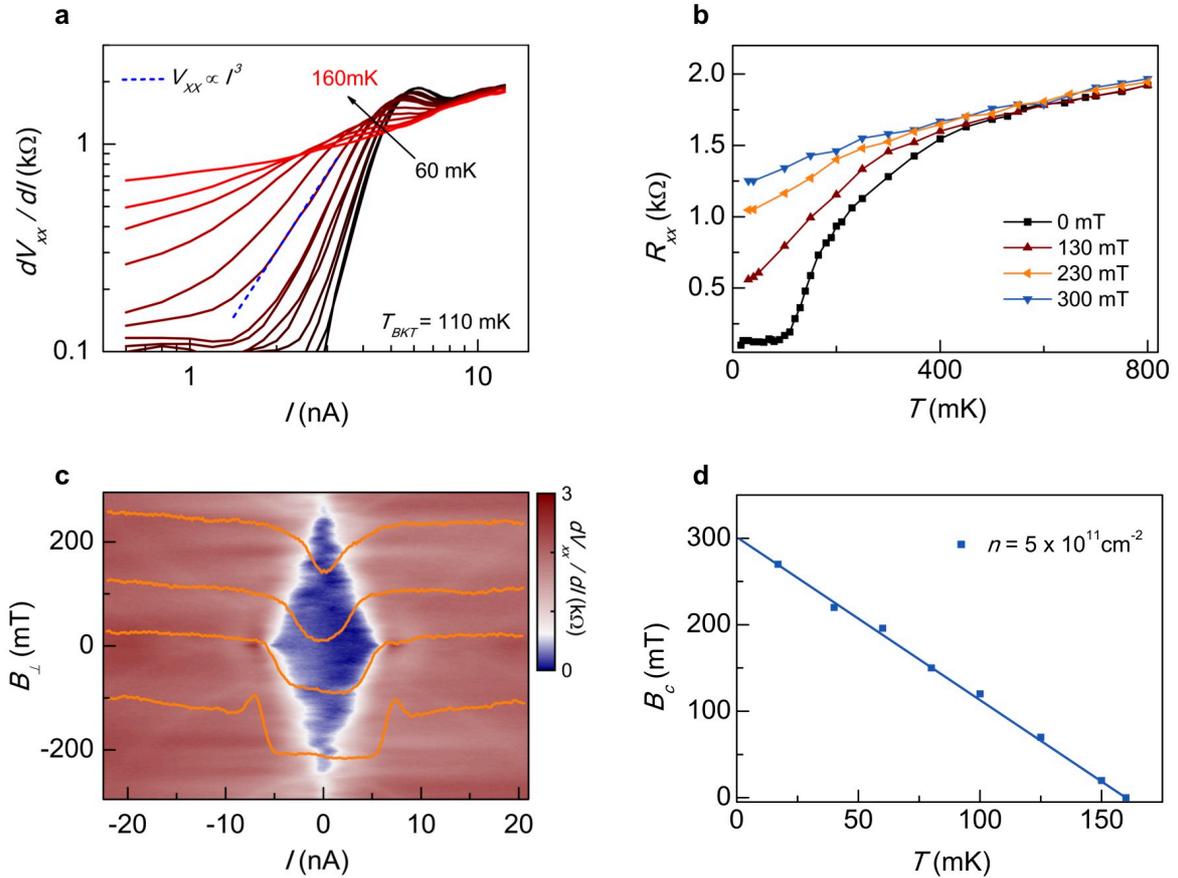

**Figure 2 | Superconducting dome between $\nu = 0$ and $\nu = 1$ filling. a,** Differential resistance $dV_{xx}/dI$ versus dc bias current $I$ at various temperatures from 60 mK (black trace) to 160 mK (red trace). The blue dashed line is a fit to the $V_{xx} \sim I^3$ power law, identifies a Berezinskii–Kosterlitz–Thouless transition (BKT) at temperature of $T_{BKT} \sim 110$ mK. **b,** Longitudinal resistance $R_{xx}$ versus temperature at various out-of-plane magnetic fields $B_\perp$, showing that normal resistance values are restored above 300 mT. **c,** Two dimensional color plot of the differential resistance $dV_{xx}/dI$ as a function of $B$ and excitation current $I$ at 16mK. The orange traces show differential resistance $dV_{xx}/dI$ versus $I$ for values of $B_\perp$ = 225 mT, 150 mT, 75 mT and 0 T from top to bottom. **d,** Critical magnetic field $B_c$ values extracted at various temperatures. The straight line is a fit to the Ginzburg-Landau theory expression. (For all of the above measurements the carrier density was fixed at optimal doping of the dome $n = 5 \times 10^{11} \text{cm}^{-2}$).

In Figure 2 we illustrate the superconducting signatures at optimal doping of the newly observed superconducting domes, exemplified by the state between the CNP and $n_0$ at $n = 5 \times 10^{11} cm^{-2}$ (all other superconducting regions are described in detail in the Extended data Figs. 6-7). Fig. 2a displays measurements of $dV_{xx}/dI$ versus dc bias current $I$ at various temperatures. At 60 mK the $dV_{xx}/dI$(I) traces display the non-linear resistance typical of 2D superconductivity, with close to zero resistance values for $I < I_c \sim 3$ nA (where $I_c$ is the critical supercurrent) and normal resistance values for $I > I_c$. The blue dashed line is a power law fit to $V_{xx} \sim I^3$, consistent with 2D superconductivity described by the Berezinskii–Kosterlitz–Thouless theory (BKT)[48]. The extracted transition temperature $T_{BKT} \sim 110$ mK is comparable with the extracted $T_c \sim 160$ mK.

The temperature dependence of the resistance $R_{xx}$(T) for various magnetic field values $B_\perp$ is illustrated in Fig. 2b. The superconductivity signal is gradually weakened by $B_\perp$, and $R_{xx}$(T) varies almost linearly with $T$ above a critical field $B_c \sim 300$ mT. The suppression of superconductivity by $B_\perp$ is further exemplified in Fig. 2c which shows a color plot of the differential resistance $dV_{xx}/dI$ as a function of $B_\perp$ and excitation current $I$ at 16 mK. The orange traces show the non-linear resistance dependence for various $B_\perp$. Here blue regions correspond to $I < I_c$, showing that $I_c$ is reduced by $B_\perp$, reaching zero above $B_c \sim 300$ mT. Unlike previous studies, our data does not show the phase coherent Fraunhofer interference patterns that are produced by phase separation between normal and superconducting regions, and have been attributed to twist-angle inhomogeneity. Their absence supports the high twist-angle homogeneity of our device and establishes the formation of a macroscopic superconducting phase. From these measurements we extract the temperature dependent critical magnetic field $B_c$ (defined by 50% of the normal state $R_{xx}$ value). By fitting to the Ginzburg-Landau theory[48] expression, $B_c = [\Phi_0/(2\pi\xi^2)](1 - T/T_c)$, we extract a coherence length $\xi_{GL(T=0K)} \sim 32$ nm. Here $\Phi_0 = h/(2e)$ is the superconducting flux quantum and $h$ is Planck's constant.

## Landau levels and Chern insulators at $\nu = \pm 1$ filling

We have studied the $B_\perp$-field response of the entire flat band at a temperature of 100 mK, identifying Shubnikov de Haas (SdH) oscillations that provide information on the structure and degeneracies of the Fermi surfaces in the sample. Fig. 3a shows the color map of $R_{xx}$ as a function of $n$ and $B_\perp$, and the corresponding schematic highlights resistance maxima trajectories. We find sets of Landau fans which originate from the CNP and from most of the integer $\nu$ resistive states. In previous work Landau levels (LL) were identified only on the high carrier density sides of insulating states. Here we also observe Landau levels (LL) dispersing to lower densities. The vanishing carrier densities near most integer $\nu$ signaled by both Landau fans and weak field Hall resistivities (Extended data Fig. 1) suggests that the four-fold spin/valley band degeneracy of the non-interacting state is lifted over large ranges of filling factor, resetting the carrier density per band.

Our observations suggest that a rich variety of spin/valley broken-symmetry states occur as a function of carrier density and magnetic field. The Landau levels (LL) which can be traced to the CNP exhibit four-fold degeneracy with a filling-factor sequence of $\nu_L = \pm 4, \pm 8, \pm 12$ etc, as well as spin/valley broken-symmetry states with $\nu_L = \pm 2$. The LLs fanning out from $\nu = 2 (-2)$, follow a sequence of $\nu_L = 2 (-2), 4 (-4), 6 (-6)$ etc. at low magnetic field indicating partially lifted degeneracy for either spin or valley. At high magnetic field, quantum oscillations from $\nu = -2$ exhibit a dominant degeneracy sequence of $\nu_L = -3, -5, -7$ etc. Near $\nu = -3$ filling, quantum oscillations exhibit fully lifted degeneracy of LLs with filling factors $\nu_L = -1, -2, -3, -4$ etc. The Landau fans that emerge from insulating states all extrapolate to a carrier density that vanishes at integer moiré band filling factors. While some details are different or newly visible here, our observations are generally in good agreement with previous reports.

We also find that the LL from the CNP and $\nu = -2$ seem to change their degeneracies, when they cross the $\nu = -1$ and $\nu = -3$ states, suggesting first order phase transitions which change band degeneracies. In particular, as is shown in Fig. 3b, the LLs from the $\nu = -2$ and $\nu = -3$ state display a crisscrossing pattern, superficially similar to that of a Hofstadter butterfly, but distinct in that the LL indices that can be traced to the $\nu = -3$ state are spaced by one filling, whereas those that can be traced to the $\nu = -2$ state, are spaced by two fillings.

Interestingly neither $\nu = \pm 1$ correlated states show clear LL formation or SdH oscillations. The positions of their resistance maxima do however exhibit clear $B_\perp$-field dependent features. At $\nu = -1$ the resistance state has no slope $dn/dB$ at low field, but suddenly, above a critical field $B_\perp \sim 3.6$T develops a slope consistent with a Chern number of 1. Even more strikingly, at $\nu = 1$ the position of the resistance peak shifts to lower carrier density, with a slope consistent with a Chern number of 2. The slope in $dn/dB$ in the absence of a LL fan in the $\nu = \pm 1$ correlated states is consistent with Chern insulating states from spin and valley symmetry breaking at odd $\nu$. As discussed in the introduction and predicted by the mean-field theory, valley projected bands in insulating states can have non-zero Chern numbers that compete closely with states with zero Chern numbers. While we cannot resolve the quantized values in either $R_{xy}$ or achieve the zero resistance in $R_{xx}$ expected for a Chern insulating state, we do not do so for the other LLs in Fig. 3a either. We therefore conclude that even our improved devices are still too inhomogeneous to observe quantization over the entire device length.

More experimental and theoretical studies will be needed to clarify these anomalous features, however overall these findings suggest quasiparticle bands that adjust to a pattern of correlations that changes when it is possible to produce gaps near the Fermi level for particular carrier densities and magnetic field strengths.

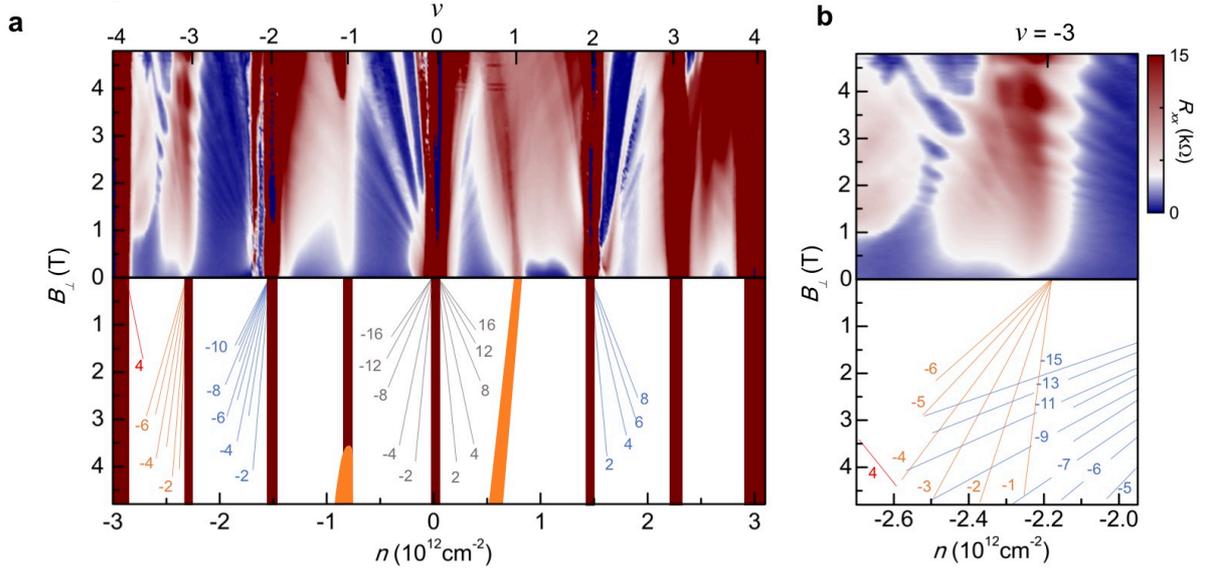

**Figure 3 | Shubnikov de Haas oscillations in the MAG flat bands. a,** (upper panel) Color map of $R_{xx}$ vs. $n$ and $B_\perp$ and (lower panel) the corresponding schematic which identifies visible Landau level fans with a dominant degeneracy. The Landau fan diagram diverging from the CNP ($\nu = 0$) follows a 4-fold degenerate sequence with $\nu_L = \pm 4, \pm 8, \pm 12$ etc., with symmetry-broken states at $\nu_L = \pm 2$. The fan from $\nu = 2 (-2)$ follows a 2-fold degenerate $\nu_L = -2 (2), -4 (4), -6 (6)$ etc. sequence, with broken-symmetry states at $\nu_L = -3, -5, -7$ etc. The $\nu = -3$ fan follows single degenerate $\nu_L = -1, -2, -3, -4$ etc. sequence. Emergent correlated phases at all integer moiré fillings, including the CNP ($\nu = 0$), are highlighted with red colors. Chern insulating states are highlighted in orange. **b,** Zoom-in of **a** around the $\nu = -3$ state, shows signatures similar to a Hofstadter butterfly spectrum with crisscrossing Landau levels fanning out from $\nu = -3$ and $\nu = -2$ filling states.

### Magnetic field driven phase transition at $\nu = -1$ filling

Surprisingly, exactly at the transition where the slope of the $\nu = -1$ resistive state in Fig. 3a changes from $dn/dB = 0$ to a $dn/dB$ consistent with Chern number 1, we find a strong hysteretic increase of $R_{xx}$. which indicates a possible magnetic field induced first-order phase transition. Fig. 4a displays a color plot of $R_{xx}$ as a function of $n$ and $B_\perp$ in which red regions correspond to high and dark blue regions to low resistance values. Fig. 4b displays the temperature-dependent resistance $R_{xx}(T)$ near $\nu = -1$ (or $n = -8.43 \times 10^{11} cm^{-2}$) for a series of magnetic field values. While at $B_\perp = 0T$, $R_{xx}(T)$ shows a typical metal-superconductor phase transition, above $B_\perp > 3.6$ T and below $T < 0.9$ K, $R_{xx}(T)$ has a sharp jump and an insulating temperature dependence.

Fig. 4c shows $R_{xx}(B_\perp)$ traces at $\nu = -1$ for up (solid) and down (dashed) sweeps of the magnetic field. Below 800 mK, the $R_{xx}(B_\perp)$ curves have sharp jumps at associated critical transition fields $B_T$, and show strong hysteretic behavior dependent on the sweeping direction of the magnetic field. The critical field $B_T$ is always higher for up sweeps than for down sweeps, and $\Delta B_T$ is the width in magnetic field of the hysteresis loop.

Both, $B_T$ and $\Delta B_T$ are highly temperature dependent, with $B_T$ shifting to higher values and $\Delta B_T$ becoming smaller as the temperature increases. For $T > 800$ mK the hysteresis almost disappears and the transition becomes broader. The temperature dependencies of both quantities are extracted and shown in Fig. 4d ($B_T$ is extracted from up sweeps). This phase transition and hysteresis occur over a narrow range of carrier density from $\sim -8.3 \times 10^{11} cm^{-2}$ to $-9 \times 10^{11} cm^{-2}$ (Extended Data Fig. 8b-c) with the transition fields $B_T$ and $\Delta B_T$ at different carrier densities shown in Fig. 4e. Combined with Fig. 4a, we see that the boundary of the transition area is quite sharp on the low doping side, where hysteresis is strongest. Overall, we observe similar behavior in Hall resistance measurements (Extended Data Fig. 8d). These observations signal that the origin of change in the slope $dn/dB$ of the resistance maximum signals a first order phase transition, and is likely due to the competition of correlated states with zero and non-zero Chern number at high magnetic fields[49], suggesting the emergence of a field stabilized orbital magnetic state.

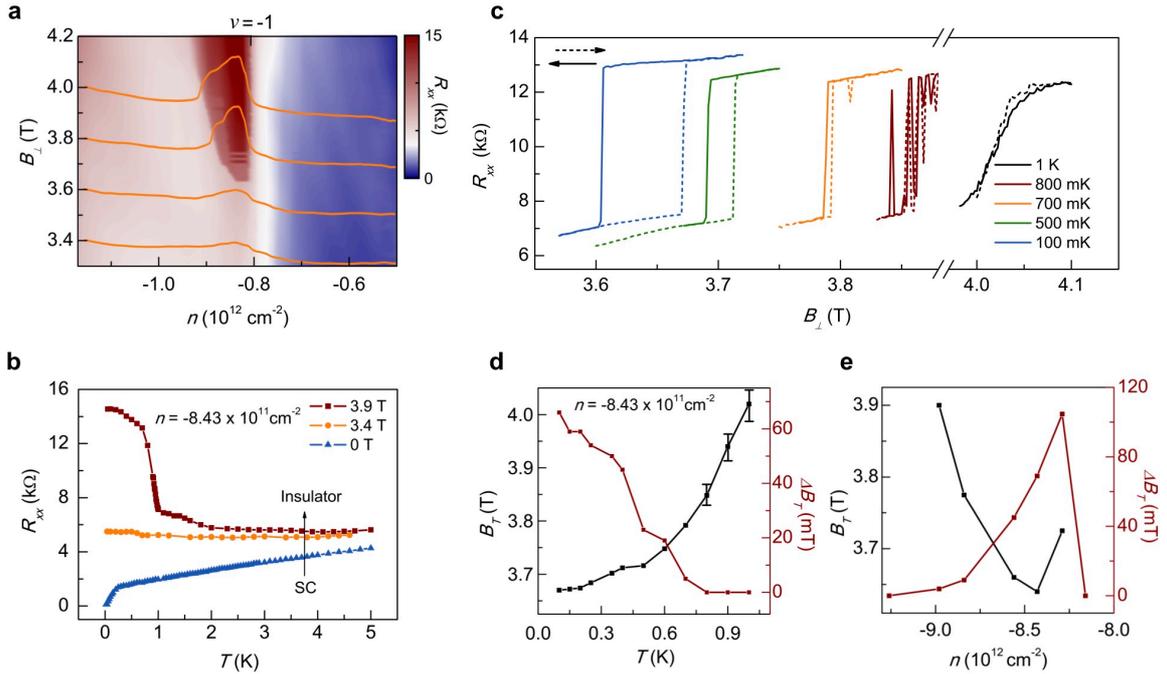

**Figure 4 | Field driven phase transition near $\nu = -1$ state. a,** Longitudinal resistance $R_{xx}$ as a function of carrier density and out-of-plane magnetic field $B_\perp$ measured at 16 mK. The orange traces show longitudinal resistance $R_{xx}$ vs. n with the magnetic field from top to bottom fixed at 4 T, 3.8 T, 3.6 T and 3.4 T, respectively. **b,** $R_{xx}$ vs temperature at various fields $B_\perp$. **c,** Magnetic field $B_\perp$ dependent resistance $R_{xx}$ at various temperatures, with dashed and solid lines corresponding to increasing and decreasing magnetic field, respectively. **d,** Temperature dependence of the critical field $B_T$ and the hysteresis value $\Delta B_T$. Note that the transition above 800 mK is not sharp, adding uncertainty to the $B_T$ extraction. (The carrier density in **b-d** is fixed at $-8.43 \times 10^{11} cm^{-2}$). **e,** Carrier density dependence of $B_T$ and $\Delta B_T$ at 100 mK.

## Discussion

By improving twist-angle homogeneity we have revealed aspects of the fascinating phenomenology of strong correlations and superconductivity in magic angle twisted bilayer graphene that were obscure in previous studies, in spite of the low impurity concentrations and relatively high mobilities of the hBN encapsulated MAG devices they employed[29,30]. The $\nu = \pm 2$ correlated states and associated superconducting domes discovered previously remain the strongest features. However, bubble-free assembly and improvements in twist-angle homogeneity in the present device have allowed a more complete picture to emerge. Most significantly, we show that insulating states occur at neutrality and at most integer moiré band filling factors, that four or more distinct superconducting domes occur as the flat band filling factor is varied, and that Chern insulators compete with normal insulators, forming a ground state at zero magnetic field in one case and stabilized by a magnetic field in another.

We observe superconducting domes close to the $\nu = \pm 1$ states, and surprisingly perhaps, also close to charge neutrality. The latter domes represent the lowest carrier density $n \sim 3 \times 10^{11} \text{cm}^{-2}$ (counting from CNP) at which SC has not ever been observed, comfortably beating previous records[29]. The existence of superconducting domes across a wide range of moiré band fillings must have important implications for our understanding of their origin. The appearance of superconductivity appears not to be simply related to proximity to peaks in the density of states of the non-interacting bands. Superconductivity occurs adjacent to insulating states that appear, on the basis of Landau fan patterns, to break spin-valley degeneracy and adjacent to insulating states that do not. Nevertheless, its consistent association with nearby correlated insulator states suggests an exotic pairing mechanism. On the other hand, our observation cannot at this point rule out the possibility of conventional electron phonon coupling superconductivity in metallic states with quasiparticles that evolve adiabatically from those of the non-interacting system and compete with a rich variety of distinct insulating states from which they are separated by first order phase transition lines[38,50,51]. In this case, it is possible that the consistent high density of states over a broad range of filling factors helps to support superconductivity in the metallic state.

Acknowledgements:

We are grateful for fruitful discussions with Pablo Jarillo-Herrero, Andrei Bernevig, Mathew Yankowitz, Andrea Young, Cory Dean, Leonid Levitov, Ashvin Vishwanath, Matthew Fisher, Milan Allan and Frank Koppens. D.K.E. acknowledges support from the Ministry of Economy and Competitiveness of Spain through the "Severo Ochoa" program for Centres of Excellence in R&D (SE5-0522), Fundació Privada Cellex, Fundació Privada Mir-Puig, the Generalitat de Catalunya through the CERCA program, the H2020 Programme under grant agreement n° 820378, Project: 2D·SIPC and the La Caixa Foundation. A.H.M. and M.X. acknowledge support from DOE grant DE-FG02-02ER45958 and Welch foundation grant TBF1473. G.Z. acknowledges supports from National Science Foundation of China under the grant No. 11834017 and 61888102, the Strategic Priority Research Program of CAS under the grant No. XDB30000000.

Author contributions:

D.K.E. and X. L. conceived and designed the experiments; X. L., W. Y. and P. S. performed the experiments; X. L. and D.K.E. analyzed the data; M. X. and A.H.M. performed the theoretical modeling of the data; T.T. and K. W. contributed materials; D. K. E., A. B., M. A. A., I. D., C. U. and G. Z. supported the experiments: X. L., D.K.E, P. S., X.M. and A.H.M. wrote the paper.

Competing financial and non-Financial interests:

The authors declare no competing financial and non-financial interests.

Data availability:

The data that support the findings of this study are available from the corresponding author upon reasonable and well-motivated request.

Methods:

Device fabrication: The hBN/tBLG/hBN/graphite stacks were exfoliated and assembled using a van der Waals assembly technique. Monolayer graphene, thin graphite and hBN flakes (~10 nm thick) were firstly exfoliated on $SiO_2$ (~300 nm)/Si substrate, followed by the "tear and stack" technique with a polycarbonate (PC)/polydimethylsiloxane (PDMS) stamp to get the final hBN/tBLG/hBN/graphite stack. The separated graphene pieces were rotated manually by the twist angle ~1.2-1.3º. We purposefully chose a larger twist angle during the heterostructure assembly due to the high risk of relaxation of the twist angle to the random lower values. To increase the structural homogeneity, we further carried out a mechanically cleaning process to squeeze the trapped blister out and release the local strain. We didn't perform subsequent high temperature annealing to avoid twist angle relaxation. We further patterned the stacks with PMMA resist and $CHF_3+O_2$ plasma and exposed the edges of graphene, which was subsequently contacted by Cr/Au (5/50 nm) metal leads using electron-beam evaporation (Cr) and thermal evaporation (Au).

Measurement: Transport measurements were carried out in a dilution refrigerator with a base temperature of 16 mK and up to 5 T perpendicular magnetic field. Standard low-frequency lock-in techniques were used to measure the resistance $R_{xx}$ and $R_{xy}$ with an excitation current of ~1 nA at a frequency of 19.111Hz. In the measurement of differential resistance dV/dI, a ~0.5 nA AC excitation current was applied through a AC signal (0.5 V) generated by the lock-in amplifier in combination with a 1/100 divider and a 10 MOhm resistor. Before combining with the excitation, the applied DC signal passed through a 1/100 divider and a 1 MOhm resistor. As-induced differential voltage was further measured at the same frequency of 19.111Hz with standard lock-in technique. For measurements in strong magnetic fields we found that the increased contact resistance made it difficult to obtain accurate device resistance values. To resolve this issue, we applied a global gate voltage (+20 V) through $Si/SiO_2$ (~300 nm) to tune the charge carrier density separately in the device leads.

---

1       Lee, P. A., Nagaosa, N. & Wen, X.-G. Doping a Mott insulator: Physics of high-temperature superconductivity. *Reviews of modern physics* **78**, 17 (2006).
2       Bistritzer, R. & MacDonald, A. H. Moiré bands in twisted double-layer graphene. *Proceedings of the National Academy of Sciences* **108**, 12233-12237 (2011).
3       Cao, Y. *et al.* Correlated insulator behaviour at half-filling in magic-angle graphene superlattices. *Nature* **556**, 80 (2018).
4       Chen, G. *et al.* Evidence of a gate-tunable Mott insulator in a trilayer graphene moiré superlattice. *Nature Physics*, 1 (2019).
5       Tarnopolsky, G., Kruchkov, A. J. & Vishwanath, A. Origin of magic angles in twisted bilayer graphene. *Physical Review Letters* **122**, 106405 (2019).
6       Yamada, A., Seki, K., Eder, R. & Ohta, Y. Mott transition and ferrimagnetism in the Hubbard model on the anisotropic kagome lattice. *Physical Review B* **83**, 195127 (2011).
7       Shen, Z.-X., Spicer, W., King, D., Dessau, D. & Wells, B. Photoemission studies of high-Tc superconductors: The superconducting gap. *Science* **267**, 343-350 (1995).
8       Kopnin, N., Heikkilä, T. & Volovik, G. High-temperature surface superconductivity in topological flat-band systems. *Physical Review B* **83**, 220503 (2011).
9       Miyahara, S., Kusuta, S. & Furukawa, N. BCS theory on a flat band lattice. *Physica C: Superconductivity* **460**, 1145-1146 (2007).
10      Sharpe, A. L. *et al.* Emergent ferromagnetism near three-quarters filling in twisted bilayer graphene. *arXiv preprint arXiv:1901.03520* (2019).

11  Xie, M. & MacDonald, A. H. On the nature of the correlated insulator states in twisted bilayer graphene. *arXiv preprint arXiv:1812.04213* (2018).
12  Ochi, M., Koshino, M. & Kuroki, K. Possible correlated insulating states in magic-angle twisted bilayer graphene under strongly competing interactions. *Physical Review B* **98**, 081102 (2018).
13  Dodaro, J. F., Kivelson, S. A., Schattner, Y., Sun, X.-Q. & Wang, C. Phases of a phenomenological model of twisted bilayer graphene. *Physical Review B* **98**, 075154 (2018).
14  Thomson, A., Chatterjee, S., Sachdev, S. & Scheurer, M. S. Triangular antiferromagnetism on the honeycomb lattice of twisted bilayer graphene. *Physical Review B* **98**, 075109 (2018).
15  Mielke, A. Ferromagnetism in the Hubbard model on line graphs and further considerations. *Journal of Physics A: Mathematical and General* **24**, 3311 (1991).
16  Tasaki, H. Ferromagnetism in the Hubbard models with degenerate single-electron ground states. *Physical review letters* **69**, 1608 (1992).
17  Nandkishore, R., Levitov, L. & Chubukov, A. Chiral superconductivity from repulsive interactions in doped graphene. *Nature Physics* **8**, 158 (2012).
18  Si, Q. & Steglich, F. Heavy fermions and quantum phase transitions. *Science* **329**, 1161-1166 (2010).
19  Mielke, A. Exact ground states for the Hubbard model on the Kagome lattice. *Journal of Physics A: Mathematical and General* **25**, 4335 (1992).
20  Mukherjee, S. *et al.* Observation of a localized flat-band state in a photonic Lieb lattice. *Physical review letters* **114**, 245504 (2015).
21  Wu, C., Bergman, D., Balents, L. & Sarma, S. D. Flat bands and Wigner crystallization in the honeycomb optical lattice. *Physical review letters* **99**, 070401 (2007).
22  Heikkilä, T. T., Kopnin, N. B. & Volovik, G. E. Flat bands in topological media. *JETP letters* **94**, 233 (2011).
23  Tang, E., Mei, J.-W. & Wen, X.-G. High-temperature fractional quantum Hall states. *Physical review letters* **106**, 236802 (2011).
24  Sun, K., Gu, Z., Katsura, H. & Sarma, S. D. Nearly flatbands with nontrivial topology. *Physical review letters* **106**, 236803 (2011).
25  Neupert, T., Santos, L., Chamon, C. & Mudry, C. Fractional quantum Hall states at zero magnetic field. *Physical review letters* **106**, 236804 (2011).
26  Cao, Y. *et al.* Strange metal in magic-angle graphene with near Planckian dissipation. *arXiv preprint arXiv:1901.03710* (2019).
27  Zou, L., Po, H. C., Vishwanath, A. & Senthil, T. Band structure of twisted bilayer graphene: Emergent symmetries, commensurate approximants, and Wannier obstructions. *Physical Review B* **98**, 085435 (2018).
28  Po, H. C., Zou, L., Senthil, T. & Vishwanath, A. Faithful Tight-binding Models and Fragile Topology of Magic-angle Bilayer Graphene. *arXiv preprint arXiv:1808.02482* (2018).
29  Cao, Y. *et al.* Unconventional superconductivity in magic-angle graphene superlattices. *Nature* **556**, 43 (2018).
30  Yankowitz, M. *et al.* Tuning superconductivity in twisted bilayer graphene. *Science*, eaav1910 (2019).
31  Cao, Y. *et al.* Superlattice-induced insulating states and valley-protected orbits in twisted bilayer graphene. *Physical review letters* **117**, 116804 (2016).
32  Song, Z. *et al.* All" Magic Angles" Are" Stable" Topological. *arXiv preprint arXiv:1807.10676* (2018).
33  Lian, B., Xie, F. & Bernevig, B. A. The Landau Level of Fragile Topology. *arXiv preprint arXiv:1811.11786* (2018).
34  Kim, K. *et al.* Tunable moiré bands and strong correlations in small-twist-angle bilayer graphene. *Proceedings of the National Academy of Sciences* **114**, 3364-3369 (2017).
35  Kim, K. *et al.* van der Waals heterostructures with high accuracy rotational alignment. *Nano letters* **16**, 1989-1995 (2016).
36  Purdie, D. *et al.* Cleaning interfaces in layered materials heterostructures. *Nature communications* **9**, 5387 (2018).
37  Wang, L. *et al.* One-dimensional electrical contact to a two-dimensional material. *Science* **342**, 614-617 (2013).
38  Polshyn, H. *et al.* Phonon scattering dominated electron transport in twisted bilayer graphene. *arXiv preprint arXiv:1902.00763* (2019).

39      Kondo, T. *et al.* Quadratic Fermi node in a 3D strongly correlated semimetal. *Nature communications* **6**, 10042 (2015).
40      Paschen, S. *et al.* Towards strongly correlated semimetals: U2Ru2Sn and Eu8Ga16Ge30. *Journal of Physics and Chemistry of Solids* **63**, 1183-1188 (2002).
41      Hunt, B. *et al.* Massive Dirac fermions and Hofstadter butterfly in a van der Waals heterostructure. *Science* **340**, 1427-1430 (2013).
42      Woods, C. *et al.* Commensurate–incommensurate transition in graphene on hexagonal boron nitride. *Nature physics* **10**, 451 (2014).
43      Wang, E. *et al.* Gaps induced by inversion symmetry breaking and second-generation Dirac cones in graphene/hexagonal boron nitride. *Nature Physics* **12**, 1111 (2016).
44      Ponomarenko, L. *et al.* Cloning of Dirac fermions in graphene superlattices. *Nature* **497**, 594 (2013).
45      Dean, C. R. *et al.* Hofstadter's butterfly and the fractal quantum Hall effect in moiré superlattices. *Nature* **497**, 598 (2013).
46      Yang, W. *et al.* Epitaxial growth of single-domain graphene on hexagonal boron nitride. *Nature materials* **12**, 792 (2013).
47      Yang, W. *et al.* Hofstadter butterfly and many-body effects in epitaxial graphene superlattice. *Nano letters* **16**, 2387-2392 (2016).
48      Tinkham, M.  Introduction to superconductivity (Courier Corporation, 1996).
49      Kagawa, F., Itou, T., Miyagawa, K. & Kanoda, K. Magnetic-field-induced Mott transition in a quasi-two-dimensional organic conductor. *Physical review letters* **93**, 127001 (2004).
50      Wu, F., Hwang, E. & Sarma, S. D. Phonon-induced giant linear-in-T resistivity in magic angle twisted bilayer graphene: Ordinary strangeness and exotic superconductivity.
51      Lian, B., Wang, Z. & Bernevig, B. A. Twisted bilayer graphene: A phonon driven superconductor. *arXiv preprint arXiv:1807.04382* (2018).

# EXTENDED DATA

## A. Hall density measurements

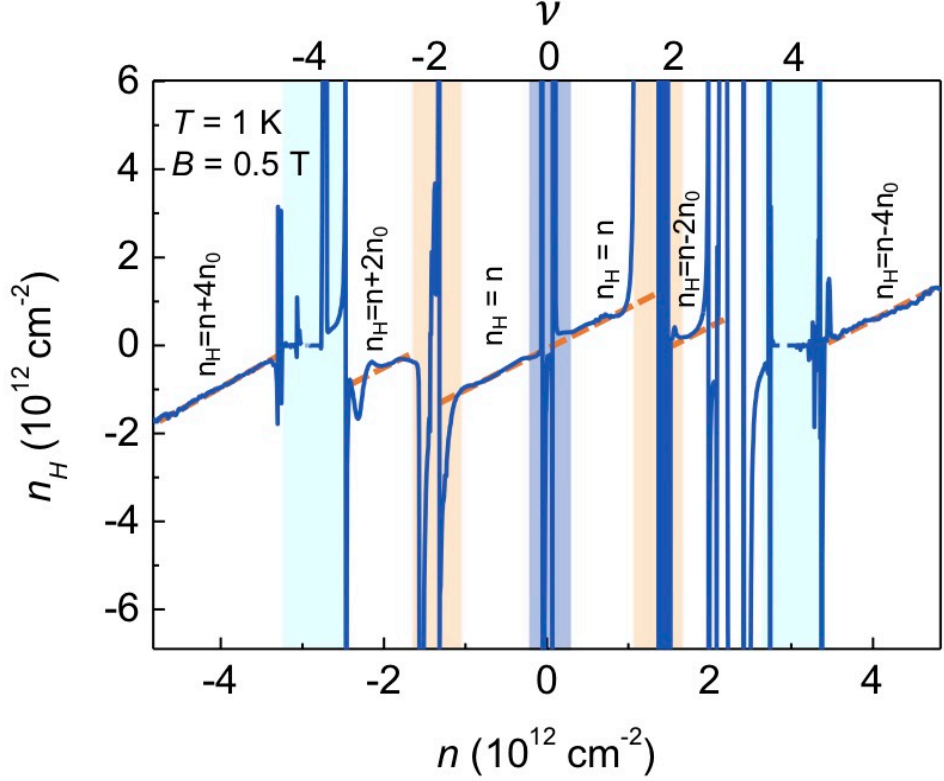

**Extended Data Figure 1 | Hall density measurements at low $B$.** Colored vertical bars correspond to densities $\nu = -4, -2, 2$ and $4$.

Extended Data Fig. 1 shows Hall charge carrier density ($n_H = -B/(eR_{xy})$) vs. total charge carrier density $n$ measured for device D1. Here we note that due to the large insulating gaps opened at the half-filled Moiré superlattice ($\nu = \pm 2$) and weaker developed features of quarter-filled states ($\nu = \pm 3, \pm 1$), the observation of charge carrier sign switching wasn't possible for each of the insulating regions. However, Extended Data Fig. 1 shows a clear charge carrier quasiparticle sign change around $n = 0$ (from $-1.3 \times 10^{12}$ cm$^{-2}$ to $1.3 \times 10^{12}$ cm$^{-2}$). As expected in a uniformly gated 2D electron gas system it follows $n_H = n$. The further increase in total charge carrier density changes $n_H = n \pm 2n_0$. We note that on the hole side the Hall density becomes negative in the region from $-1.6 \times 10^{12}$ cm$^{-2}$ to $-2.5 \times 10^{12}$ cm$^{-2}$ and positive on the electron side from $1.5 \times 10^{12}$ cm$^{-2}$ to $2.0 \times 10^{12}$ cm$^{-2}$. Beyond the band insulator regions, we observe Hall density that follows strictly $n_H = n \pm 4n_0$.

In a good agreement with quantum oscillation data on the Fig. 4a in the main text, we observe new quasiparticles that appear only on one side of the CS states. This may be explained by the effective mass change through the metal-insulator transition region, which obscures the observation of Hall effect.

## B. Twist angle homogeneity measurements

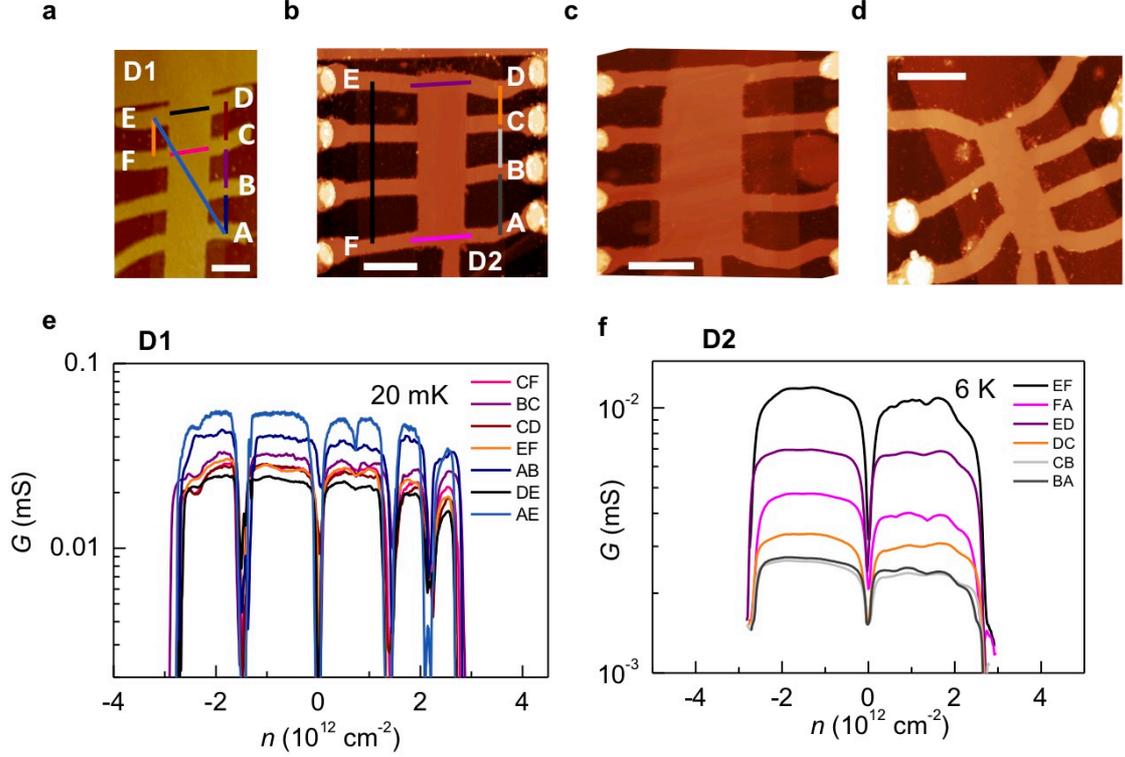

**Extended Data Figure 2 | Twist angle homogeneity measurements. a-d** AFM micrographs of a set of tBLG samples. Scale bar is 2 μm. **e-f,** Two terminal conductance measurements taken between contacts shown in a) and b). Colors correspond to the bars shown on the figures **a** and **b**, respectively.

We present extended data for two magic-angle tBLG samples D1 and D2. D1 is the sample whose data are presented in the main text. In addition, we acquired data for sample D2 at the temperatures as low as 6 K. Extended Data Fig. 2a-d show the surface topology of a set of tBLG samples. We note that all our samples exhibit low surface roughness <0.1 nm and no signatures of air bubbles that are known to locally distort carrier density in charged devices. To further check twist angle homogeneity, we find positions of the full-filled superlattice on the charge carrier density map. High-density two-terminal conductance dips correspond to the full-filled superlattice bands. Since these dips are quite wide for D1, we choose the position of $\pm 4n_0$ at the points where SdH oscillations emanating from the band insulator regions merge at $B_\perp = 0$ T. For the devices D2, we extract these values from Hall density measurements. As a result, we estimate the twist angle for D1 ~1.10° ± 0.01° and for D2 ~1.08° ± 0.01°. The two terminal conductance line cuts are shown on the Extended Data Fig. 2c (d) for D1 (D2), respectively and indicate high angle homogeneity in both samples.

## C. Magnetic field effect

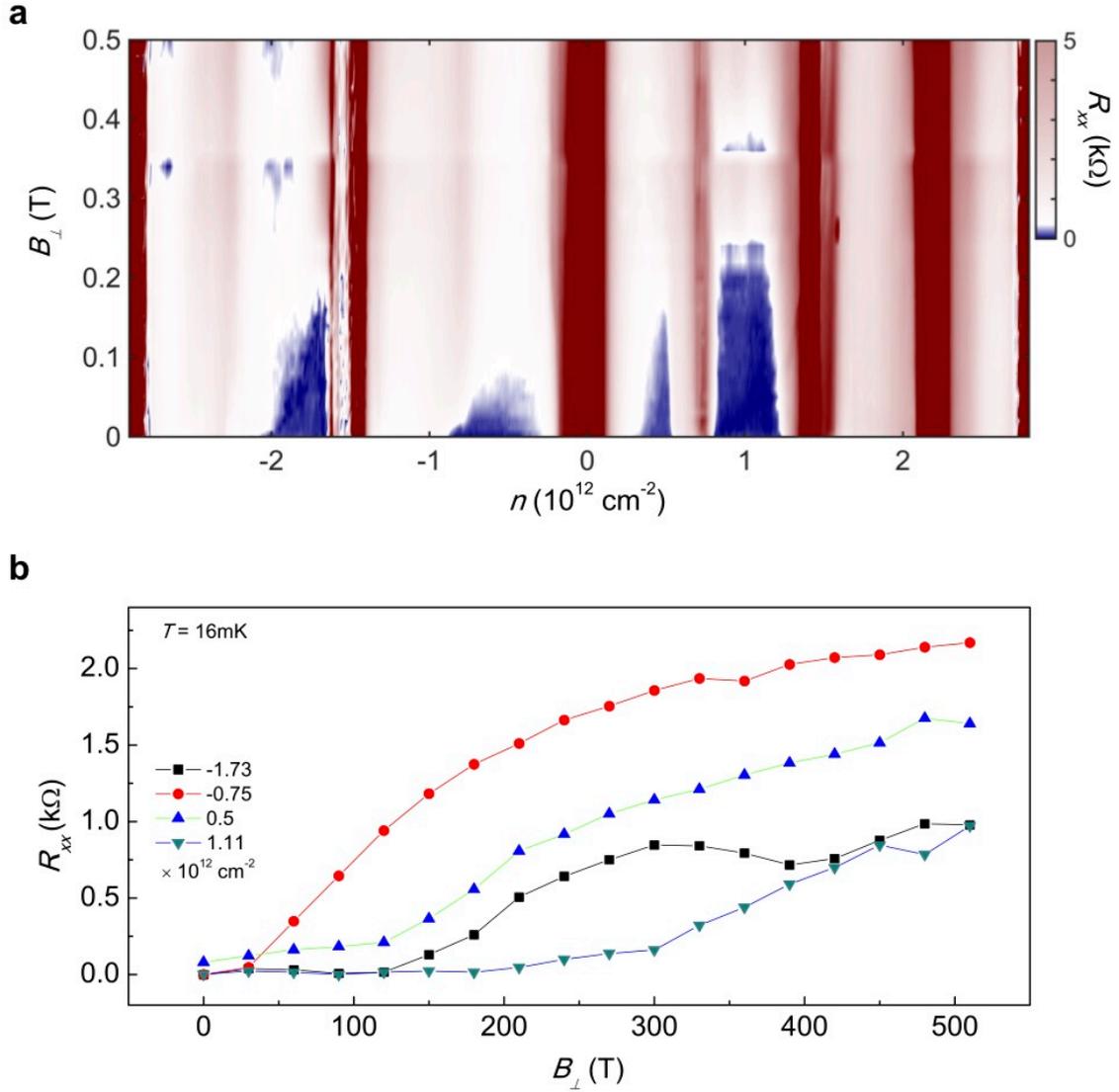

**Extended Data Figure 3 | Magnetic field measurement at base temperature. a,** 2D map $R_{xx}$ vs. $B_\perp$, $n$. The blue regions indicated SCs. **b,** Line cuts $R_{xx}$ vs. $B_\perp$ for each of the SC pockets.

Extended Data Fig. 3 demonstrates the effect of perpendicular magnetic field $B_\perp$ on the SC pockets observed in sample D1. Fig. 3a shows 2D map of longitudinal resistance $R_{xx}$ as a function of $B_\perp$ and total charge carrier density $n$ taken at the base temperature 16 mK. Extended Data Fig. 3b shows $R_{xx}$ vs. $B_\perp$ taken at 16 mK for each SC at the optimal doping level.

## D. Temperature line cuts

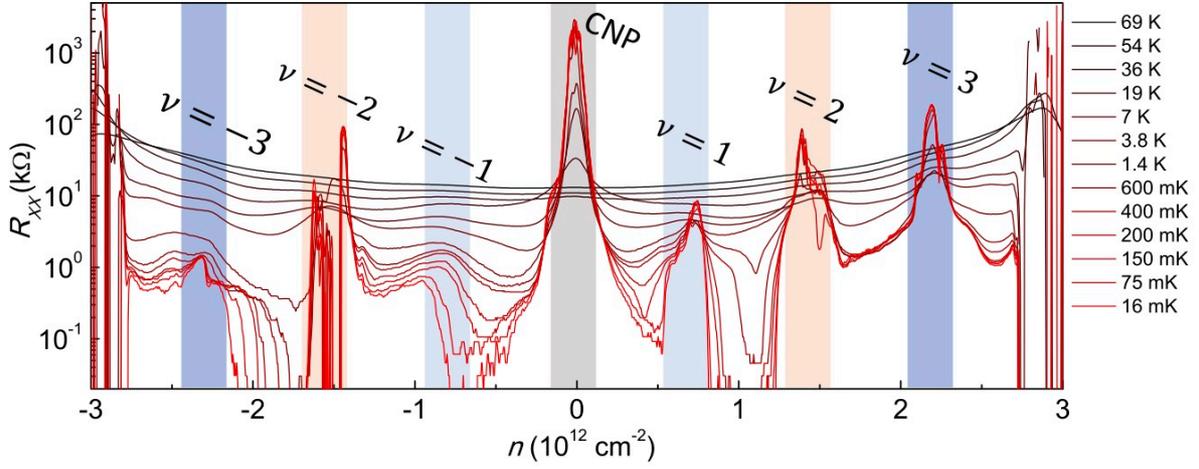

**Extended Data Figure 4 | $R_{xx}$ vs. $n$ line cuts taken at different temperatures.** Red (black) line correspond to 16 mK (69 K), respectively.

## E. Signatures of superconductivity around other correlated gapped states

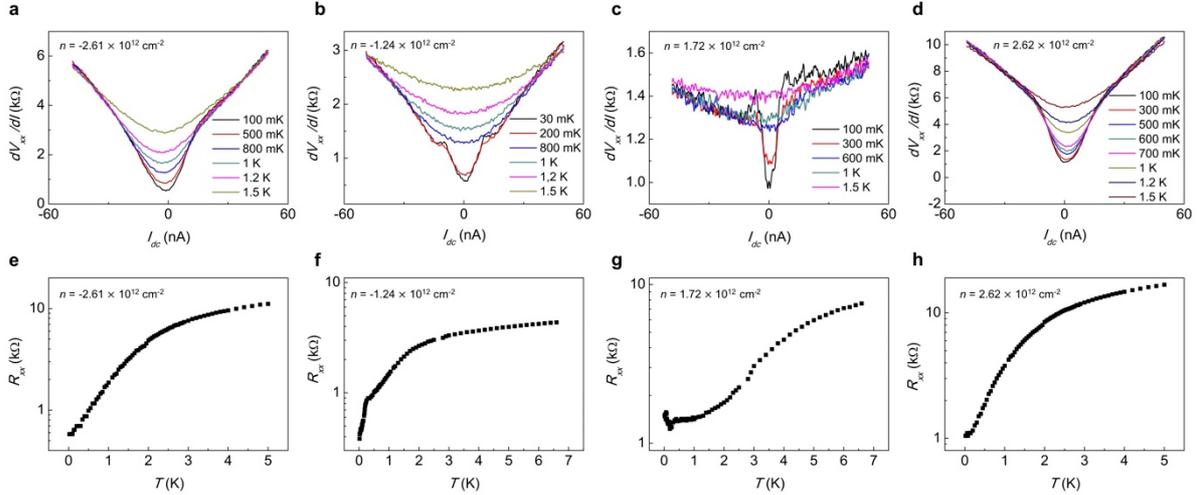

**Extended Data Figure 5 | Additional measurements at other possible superconducting domes. a-d,** $dV_{xx}/dI$ measurements for additional domes between $-4n_0$ and $-3n_0$, $-2n_0$ and $-n_0$, $2n_0$ and $3n_0$, and between $3n_0$ and $4n_0$, respectively. **e-h,** Corresponding thermal activation measurements $R_{xx}$ vs. $T$ for the same carrier densities.

Extended Data Fig. 5 shows transport measurement for charge carrier densities close to different correlated insulator regions. Extended Data Fig. 5a-d show differential resistance measurements for charge carrier densities $n = -2.61 \times 10^{12}$ cm$^{-2}$, $-1.24 \times 10^{12}$ cm$^{-2}$, $1.72 \times 10^{12}$ cm$^{-2}$, $2.62 \times 10^{12}$ cm$^{-2}$, which are between $-4n_0$ and $-3n_0$, $-2n_0$ and $-n_0$, $2n_0$ and $3n_0$, and between $3n_0$ and $4n_0$, respectively. This data suggests SC-like gap opening for the above charge carrier densities, but due to disorder or angle inhomogeneity are not fully developed. Thermal activation behavior shown on Extended Data Fig. 2e-h also suggests strong metallic behavior for the above densities, except one at $n = 1.72 \times 10^{12}$ cm$^{-2}$, at which the gap is not fully developed and start deviating from the metallic behavior at ~0.3 K. The data shown in the Extended Data Fig. 5 suggests that, possibly, one can observe four more SC pockets in a case of more homogeneous and cleaner device.

## F. Full dataset for all observed SC states

| SC pocket density | $T_c$ (mK) | $T_{BKT}$ (mK) | $B_{C\,(T=0)}$ (mT) | $\xi_{GL(T=0)}$ (nm) |
|---|---|---|---|---|
| $-1.73 \times 10^{12}$ cm$^{-2}$ | 3000 | 600 | ~180 | ~41 |
| $-7.6 \times 10^{11}$ cm$^{-2}$ | 140 | 75 | ~100 | ~55 |
| $5 \times 10^{11}$ cm$^{-2}$ | 160 | 580 | ~300 | ~32 |
| $1.11 \times 10^{12}$ cm$^{-2}$ | 650 | 110 | ~400 | ~27 |

**Extended Data Table 1.** Experimental data for all four SC states reported in this study.

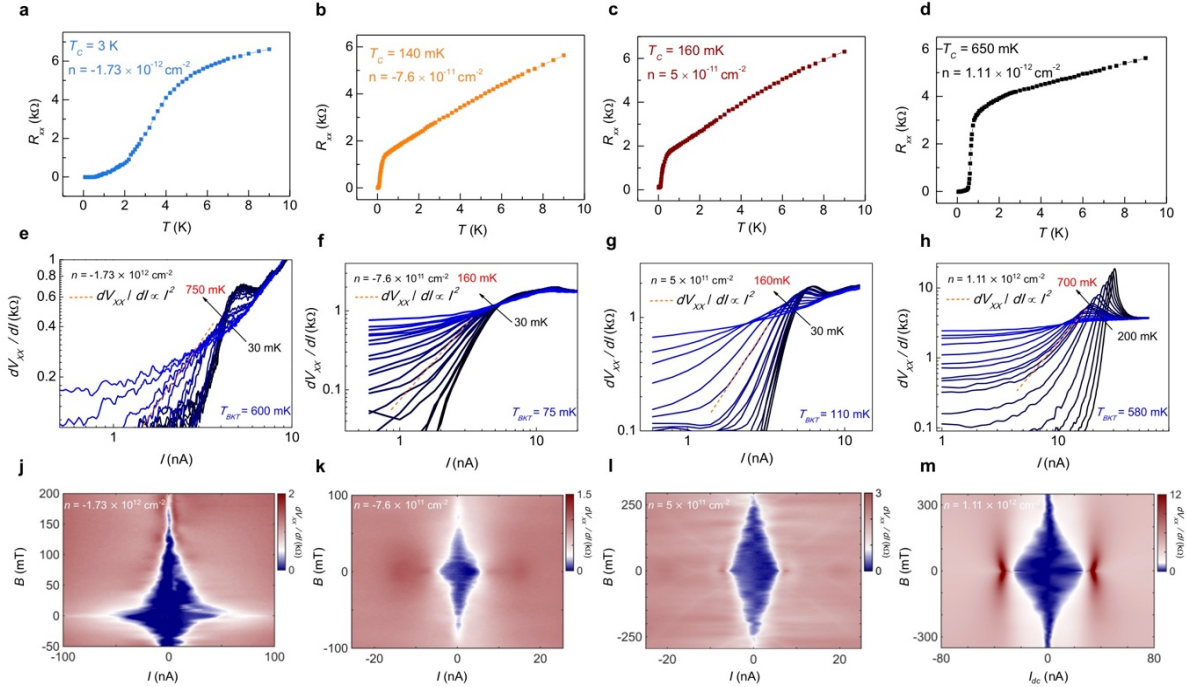

**Extended Data Figure 6 | Full characterization of all four SC pockets in sample D1**. **a-d,** thermal activation behavior curves $R_{xx}$ vs. $n$. **e-h,** BKT transition measurements $dV_{xx}/dI$ vs. $I_{dc}$. **j-m,** $R_{xx}$ vs. $B_\perp$, $I_{dc}$ maps that demonstrate the absence of Fraunhofer interference in the observed SCs.

We observed a total of four SC pockets in the reported device D1. Extended data Fig. 6 shows thermal activation (a-d), Berezinskii-Kosterlitz-Thouless (BKT) transition (e-h) and critical current, critical magnetic field measurements for all the SCs reported in D1. All SCs demonstrate strong thermal activation features with critical temperatures varying from 140 mK to 3 K. BKT transitions also demonstrate strong activation behavior with $T_{BKT}$ varying from 75 mK to 600 mK.

It is known that in the presence of strong magnetic fields 2D SC gaps become less robust due to the vortices that introduce dissipation and suppress superconductivity. Extended Data Fig. 6j-m demonstrate all critical magnetic field and current maps $dV_{xx}/dI$ vs. $B_\perp$, $I_{dc}$. These figures show well developed "diamond" – like features for all pockets with no signatures of Fraunhofer interference oscillations. The absence of such oscillations is an indicator of a well-developed SC across the entire area between the measurement contacts. This also highlights high angle and charge density homogeneity of the reported sample.

### G. Zoomed in data for phase diagram around $\nu = -1$, CNP and $\nu = 1$

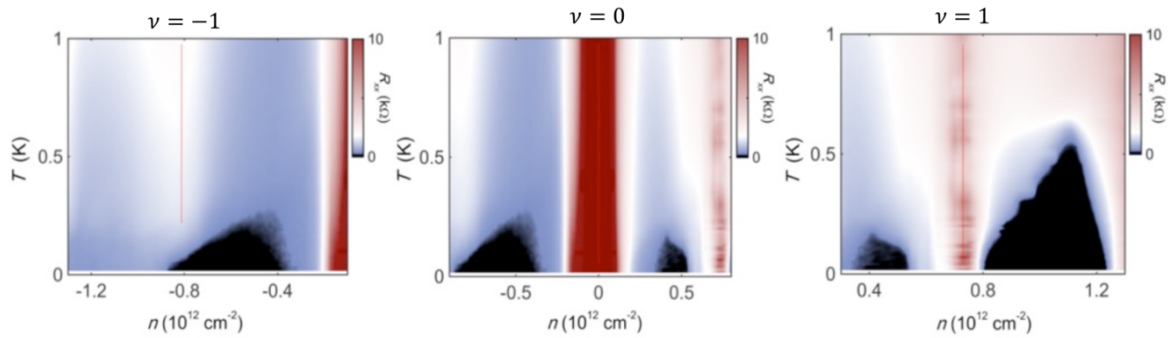

**Extended Data Figure 7 | Zoomed-in images of superconductor correlated insulator phases reported in Fig.1e in the main text.** Black regions indicate resistance below 100 Ohm. Red dashed lines indicate correlated gapped states. SC-CS phase diagram observed close to **a,** $\nu = -1$. **b,** CNP and **c,** $\nu = 1$.

## H. Additional data on hysteresis around $\nu = -1$

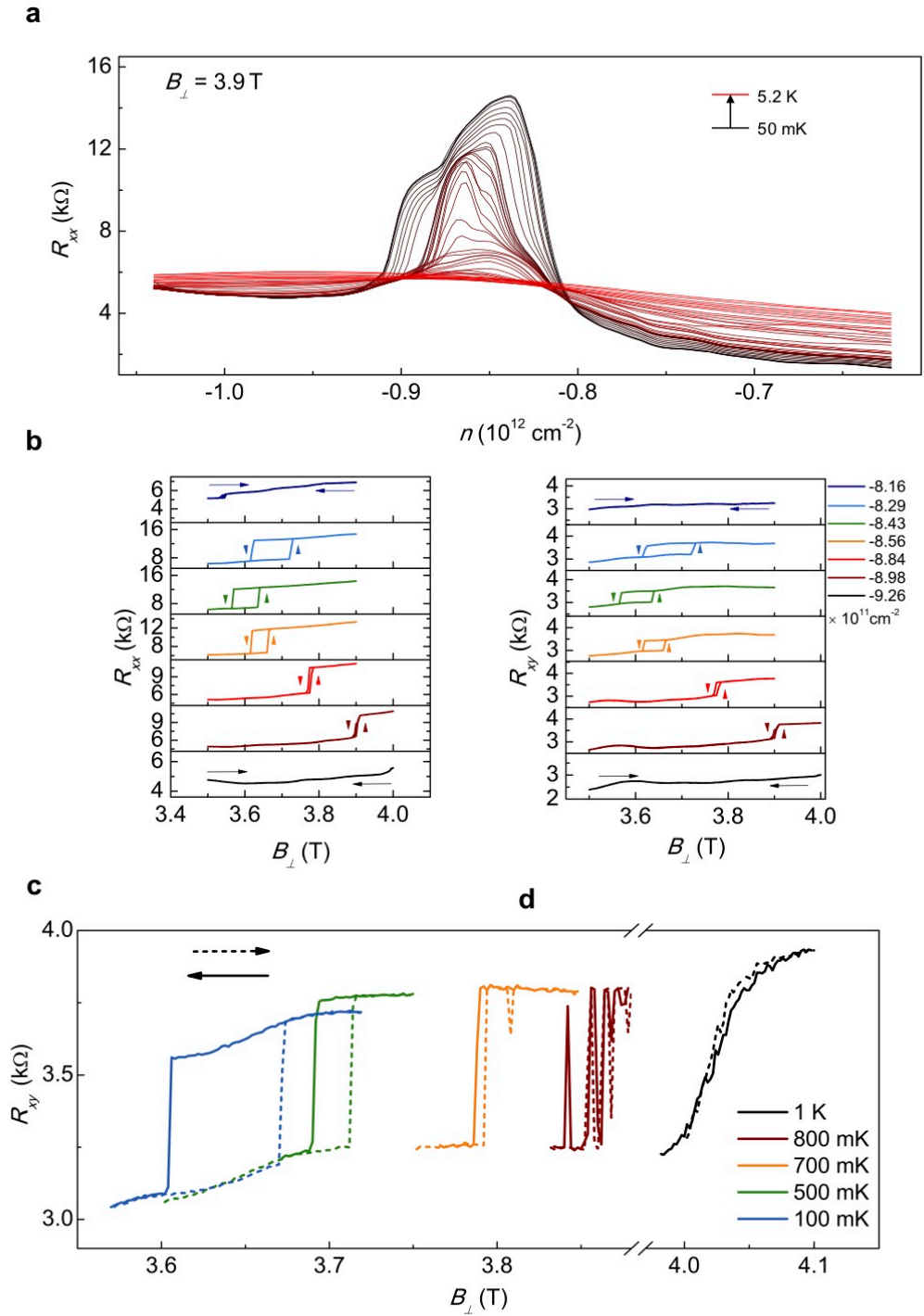

**Extended Data Figure 8 | Additional magnetic hysteresis data. a,** $R_{xx}$ vs. $n$ line cuts taken at different temperatures and used to extract data for Fig. 3b in the main text. **b-c,** $R_{xx}$ and $R_{xy}$ vs. $B_\perp$ curves taken at different charge carrier densities and 100 mK. Arrows indicate the sweep direction of the magnetic field. Data from **b** is used to extract data for Fig. 4e in the main text. **d,** Same dataset for $R_{xy}$ vs. $B_\perp$ as on the Fig. 4c in the main text. Dashed and solid lines correspond to ascending and descending magnetic fields, respectively.

## I. Quantum oscillations at high densities

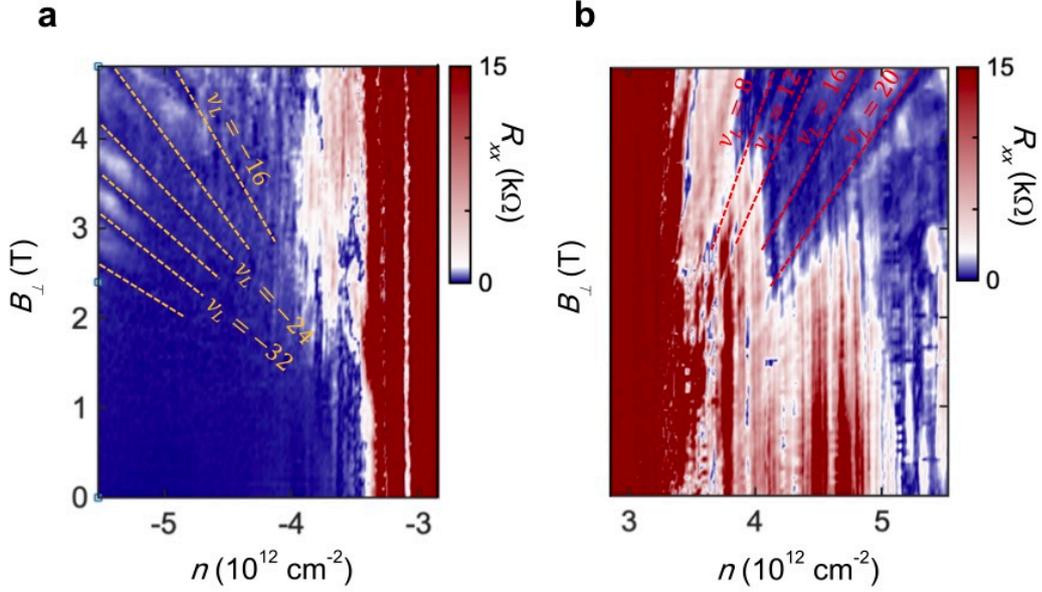

**Extended Data Figure 9 | SdH oscillation at high magnetic fields and high charge carrier densities beyond the band edges.** SdH oscillations for electron (hole) sides are shown on **a (b)**.

Extended Data Fig. 9 a-b show $R_{xx}$ vs. $B_\perp$ and $n$ taken at high charge carrier densities and high magnetic fields. The figure shows SdH quantum oscillations that appear outside the full-filled Moiré superlattice unit cell. We note that at high charge carrier densities the LL fan diagram follows that of Bernal stacked bilayer graphene. We observe a sequence of resistance dips corresponding to hole-like LL with integer fillings $\nu_L$ = 16, 20, 24, 28, 32, 36 (Extended Data Fig. 9a) and electron-like LL with integer fillings $\nu_L$ = 8, 12, 16, 20 (Extended Data Fig. 9b).

## J. SC-correlated state relation

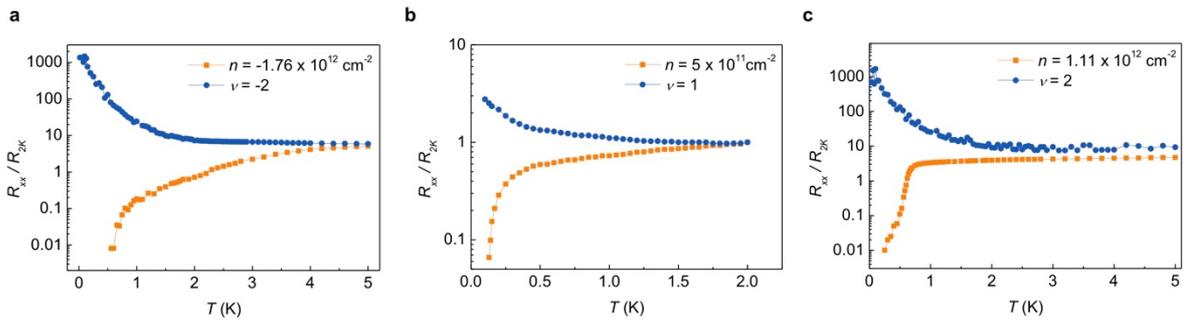

**Extended Data Figure 10 | SC – correlated state relations. a-c,** $R_{xx}$ vs. $T$ for different SC pockets and corresponding correlated insulators.

Extended Data Fig. 10a-c show the relation between correlated state and corresponding superconductivity. We note that the two phases onset at very similar temperatures with the resistance in both cases diverging from high temperature metallic behavior at 1 K (Fig. 10a), 0.5 K (Fig. 10b), and 0.7 K (Fig. 10c). This observation suggest that correlated states and superconductor gaps obey a very similar energy scale.

## K. High temperature data

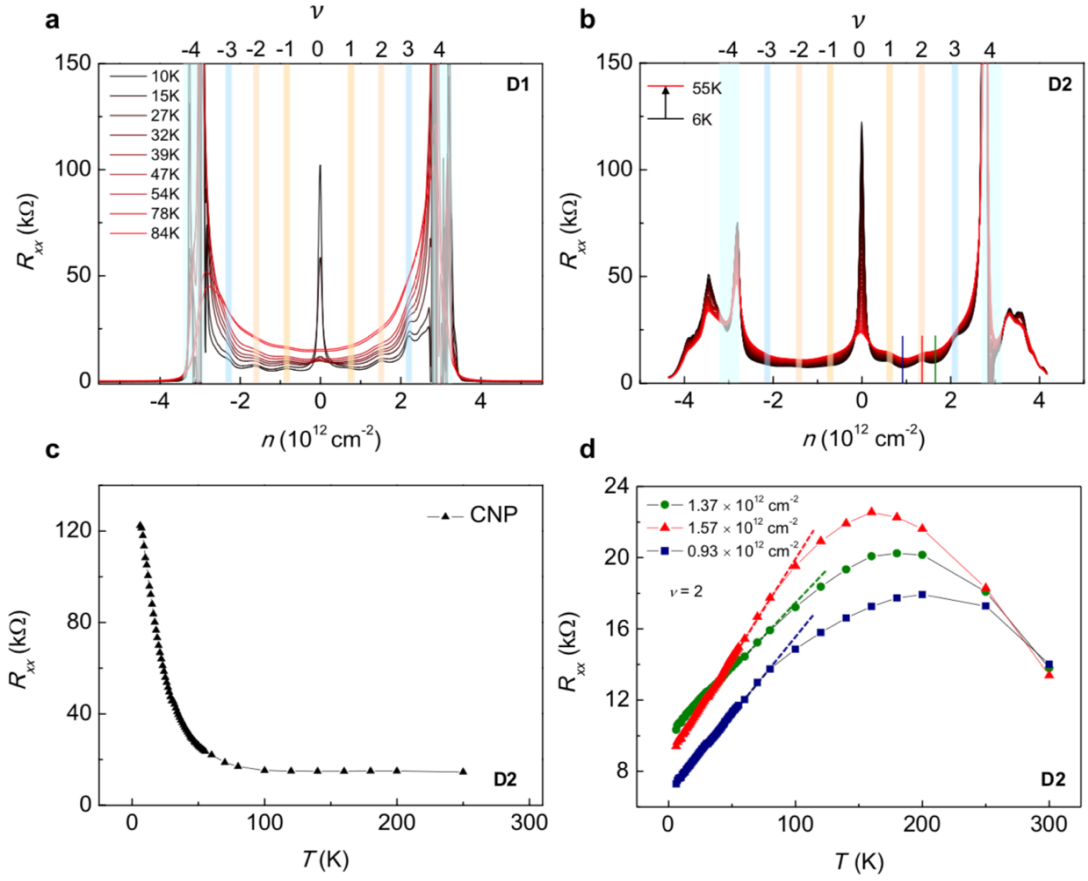

**Extended Data Figure 11 | High temperature transport measurements for Devices D1 and D2. a-b,** $R_{xx}$ vs. $n$ for samples D1 and D2, respectively. **c,** CNP thermal activation behavior for device D2. **d,** $R_{xx}$ vs. $n$ line cuts for device D2 at charge carrier densities taken along lines indicated in b (color-coded).

For the general purpose of characterization Extended Data Fig. 11a-b demonstrate $R_{xx}$ vs. $n$ scans taken at high temperatures for both samples D1 and D2. We observe the emergence of CNP at ~30 K for sample D1 and at ~90 K for sample D2. We also observe the appearance of local resistance maxima at the integer fillings of superlattice unit cell for both samples at ~50 K. Extended Data Fig. 11c shows thermal activation behavior for CNP of sample D2 from 6 K up to 250 K. Extended Data Fig. 11d shows metal-like temperature dependencies for device D2 at the fillings shown on the Extended Data Fig. 11b. Dashed lines demonstrate regions with the close-to-linear behavior.

## L. Quasiparticle bandstructures of insulating ground states at $\nu = 0$

We performed self-consistent mean field calculations, adding to the Coulomb interactions starting from the non-interacting continuum model for twisted bilayer graphene[1]. Details of the calculation can be found in Ref. 2. One important parameter in our calculation is the ratio of intra-sublattice to inter-sublattice tunneling amplitudes $T_{AA}/T_{AB}$ which controls the bandgap between flat bands and higher energy bands. Its value can be adjusted to account reasonably accurately for corrugation and lattice relaxation effects, which may be somewhat sample dependent. We use $T_{AA}/T_{AB} = 0.8$, appealing to the thorough DFT modelling study of Ref. 3.

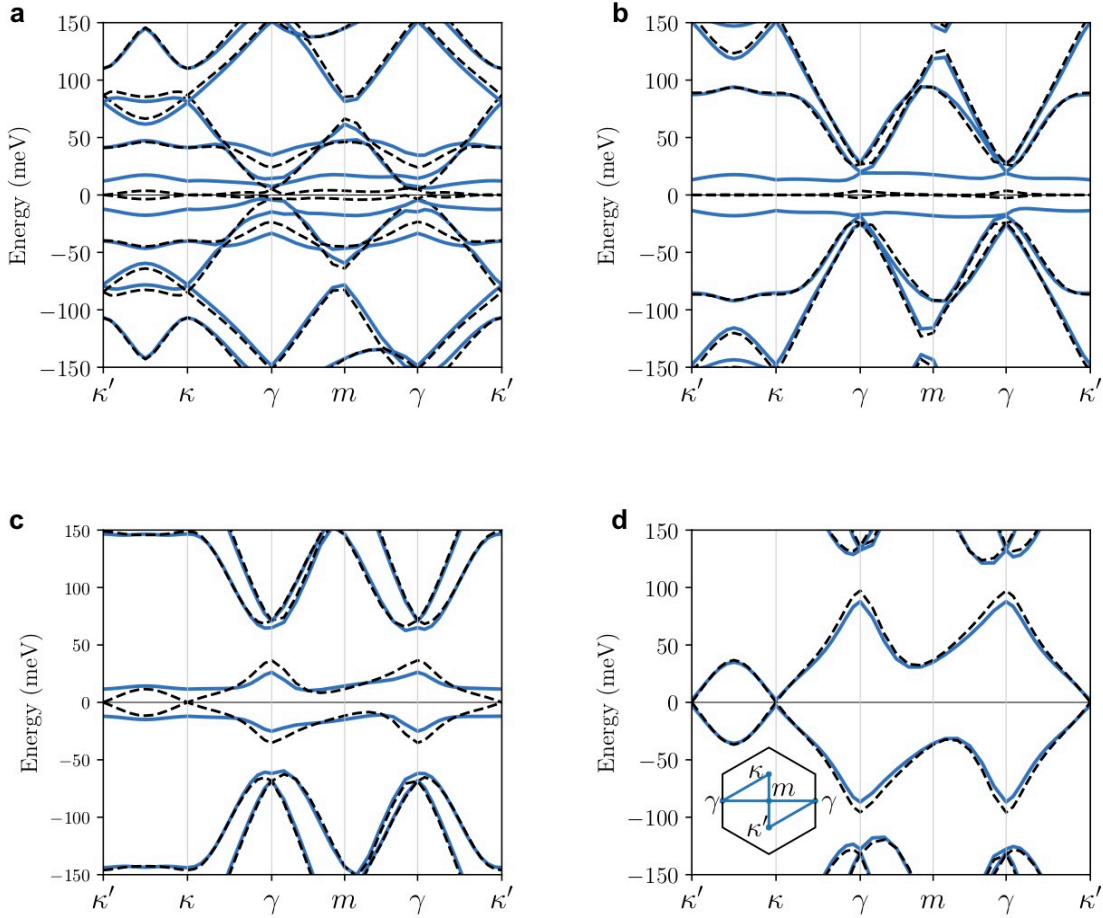

**Extended Data Figure 12| Quasiparticle bandstructures obtained by performing self-consistent mean field ground states at neutrality ($\nu = 0$) with twist angle a,** θ=0.85º, **b,** θ=1.10, **c,** θ=1.40º and **d,** θ=1.80º. The interaction strength parameter used for all this calculations was $\varepsilon^{-1} = 0.06$. The dashed lines illustrate the non-interacting bandstructure as comparisons. Case **b** is close to the magic angle condition. The inset of **d** shows the moiré Brillouin zone and its high symmetry points. These plots are results of single flavor calculation, which differs only slightly from four flavor calculations at neutrality.

Extended Data Fig. 12 shows the mean field quasiparticle bandstructures for different phases across the phase diagram Fig. 1a. We fixed the interaction strength to be $\varepsilon^{-1} = 0.06$ as an example. The four twist angles fall into three distinct ground states as classified in Fig 1g. Extended Data Fig. 12a and Fig. 12c correspond to states that break $C_2T$ symmetry. While the highest valence band of Fig. 12a has a nonzero Chern number, it vanishes for Fig.12c. Fig. 12 b and Fig. 12d correspond to states which do not break $C_2T$ symmetry but still have a finite charge gap. The mechanism of gap opening in this case is unusual and involves the annihilation of the flat band Dirac points by Dirac points sourced by band crossings between flat bands and remote higher energy bands[2]. It is evident in the Fig. 12a and Fig. 12b that the originally isolated flat bands get "pushed" into the higher energy bands by interactions and have band touching points with them away from high symmetry momenta. The difference between non-zero Chern number (Fig. 12a) and zero Chern number (Fig. 12c) bands is related to band touchings with the higher energy bands. We remark that the interaction strength in (Fig. 12c) is weak in comparison to the band gap between flat bands and higher bands. The dominant interaction effects shown in these plots is renormalization of the lowest bands only.

## M. Quasiparticle bandstructure of an insulating ground state at negative quarter filling ($\nu = -1$)

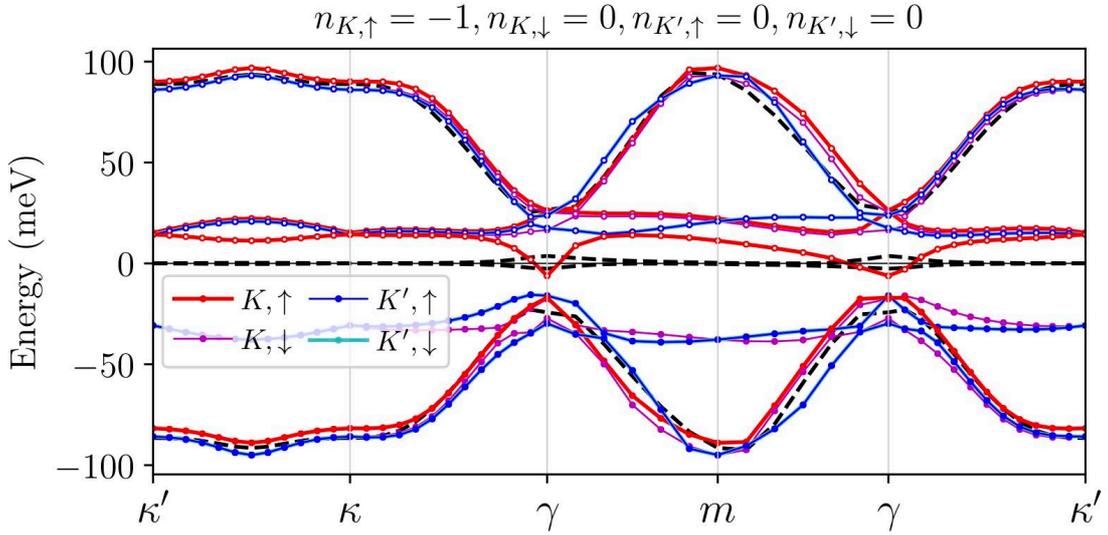

**Extended Data Figure 13 | Four-flavor quasiparticle bandstructures for an insulating ground states at $\nu = -1$ at magic angle θ=1.10º.** We chose $\varepsilon^{-1} = 0.06$ as an example. The black dashed lines are the non-interacting bandstructures, which are shown for comparison. Filled and empty circle markers represent occupied and unoccupied bands, respectively. We plot the lowest four bands, i.e. two conduction and two valence bands, for each flavor. $n_{K(K'),\uparrow(\downarrow)}$ is defined in unit of $n_0$. The ground state spontaneously polarizes to $|K,\uparrow\rangle$ flavor.

Extended Data Fig. 13 shows the quasiparticle bandstructure for filling factor $\nu = -1$. The ground state spontaneously breaks the SU(4) flavor symmetry by emptying the highest valence band of $|K,\uparrow\rangle$ flavor. This emptied valence band is shifted up and separated from the quasi-degenerate valence bands of the other flavors. Depending on the detailed parameters, the ground state may either be insulating with a small gap or semi-metallic with a small band overlap. For this value of $\varepsilon^{-1}$ the ground state is semi-metallic with a small band overlap, which occurs not along the high symmetry lines and is therefore not visible in this plot. The bandstructures at other filling factors can be understood in a similar way[2].

**References**

1. R. Bistritzer and A. H. MacDonald, Moiré bands in twisted double-layer graphene, *Proc. Natl. Acad. Sci. U.S.A.* **108**, 12233 (2011).
2. M. Xie and A. H. MacDonald, On the nature of the correlated insulator states in twisted bilayer graphene, *arXiv: 1812.04213*.
3. S. Carr, S. Fang, Z. Zhu, and E. Kaxiras, Minimal model for low-energy electronic states of twisted bilayer graphene, *arXiv:1901.03420.*